\documentclass[english,utf8,biblatex]{lni}
\usepackage{booktabs}
\usepackage{siunitx}
\begin{document}

\title[Radio Galaxy Classification with wGAN-Supported Augmentation]{Radio Galaxy Classification with wGAN-Supported Augmentation}

\author[J.~Kummer et al.]{
  Janis Kummer\footnote{Center for Data and Computing in Natural Sciences (CDCS), Notkestrasse 9, D-22607 Hamburg, Germany}$^{,}$\footnote{Universit\"at Hamburg, Hamburger Sternwarte, Gojenbergsweg 112, D-21029 Hamburg, Germany \\ \email{janis.kummer@uni-hamburg.de}}$\ $ \and
  Lennart Rustige$^{1,}$\footnote{Deutsches Elektronen-Synchrotron DESY, Notkestrasse 85, D-22607 Hamburg, Germany \\ \email{lennart.rustige@desy.de}}$\ $ \and
  Florian Griese$^{1,}$\footnote{Section for Biomedical Imaging, University Medical Center Hamburg-Eppendorf, D-20246 Hamburg, Germany \\ Institute for Biomedical Imaging, Hamburg University of Technology, D-21073 Hamburg, Germany \\ \email{florian.griese@tuhh.de}}$\ $
 \and 
  Kerstin Borras$^{3,}$\footnote{RWTH Aachen University, Templergraben 55, D-52062 Aachen, Germany}$\ $
  \and
 Marcus~Br\"uggen$^{2}\ $
 \and 
  Patrick L. S. Connor$^{1,}$\footnote{Universit\"at Hamburg, Institut für Experimentalphysik, Luruper Chaussee 149, D-22761
Hamburg, Germany\\
 }$\ $
 \and
 Frank~Gaede$^{3}\ $
 \and
 Gregor~Kasieczka$^{6}\ $
 \and 
 Peter~Schleper$^{6}\ $
}

\startpage{469}
\editor{D. Demmler, D. Krupka, H. Federrat}
\booktitle{INFORMATIK 2022 Workshops}
\yearofpublication{2022}
\lnidoi{10.18420/inf2022\_38}
\maketitle

\begin{abstract}
Novel techniques are indispensable to process the flood of data from the new generation of radio telescopes. In particular, the classification of astronomical sources in images is challenging. Morphological classification of radio galaxies could be automated with deep learning models that require large sets of labelled training data. Here, we demonstrate the use of generative models, specifically Wasserstein GANs (wGAN), to generate artificial data for different classes of radio galaxies. 
Subsequently, we augment the training data with images from our wGAN.  We find that a simple fully-connected neural network for classification can be improved significantly by including generated images into the training set. 
\end{abstract}


\section{Introduction}

The new generation of radio telescopes (e.g. LOFAR, MeerKAT and in the future the SKA \cite{LOFAR, MeerKAT, Carilli_2004}) will produce enormous amounts of data and the improved sensitivity of the instruments leads to a much higher source density within these data sets. As a result, a new level of automation for processing the data and for classifying sources is needed. Morphological classification of radio sources is crucial for answering a range of fundamental astrophysical questions, such as the origin of cosmic magnetism. 
One promising approach for morphological classification relies on the use of deep classifiers trained on well-understood data sets. However, the existing amount of data with morphological labels is limited, as they are typically extracted from catalogues created and curated manually by experts. 
Small data sets used in the training of deep learning models for galaxy classification can be enlarged by data augmentation, e.g. by applying random rotations and reflections to the images (classical augmentation). In this work, we investigate a novel application of generative models to enhance the available training sets. For this augmentation technique, multiple neural networks are combined to learn the underlying distribution of a data set. This allows the creation of new data by sampling from the learnt distribution. 
In this study we investigate whether radio galaxy classifiers can be improved when trained on such enhanced data sets containing generated data. For similar approaches from different fields see for example \cite{Frid-Adar_2018,Zhu_2017}.

\begin{figure*}
    \centering
    \includegraphics[width=0.9\textwidth]{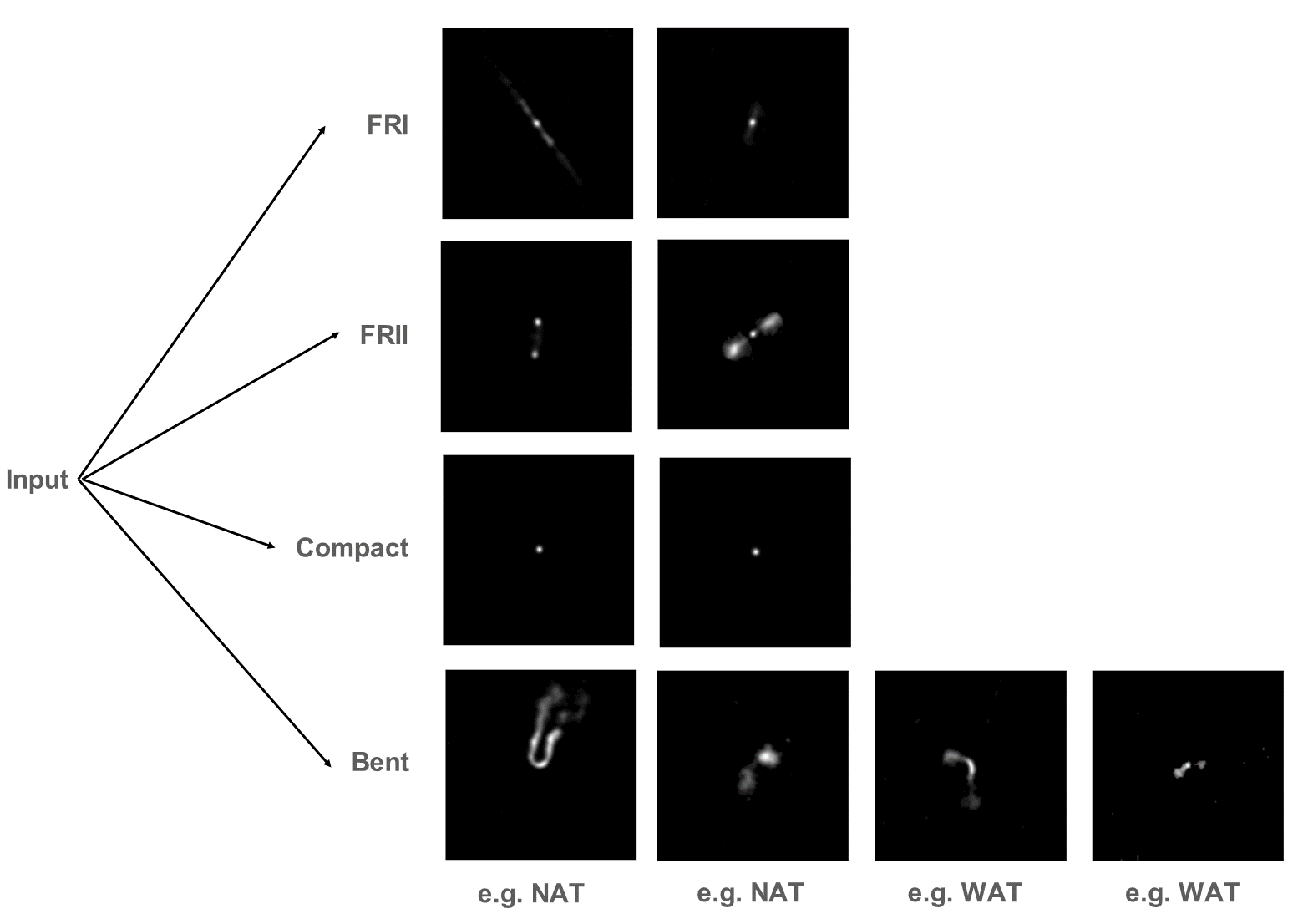}
    \caption{Examples of the classes FRI, FRII, Compact and Bent.}
    \label{fig:my_label}
\end{figure*}

As radio sources exhibit a large variety of structures, we consider a four-class classification problem (as in \cite{Alhassan_2018,Samudre_2021}), including bent-tail and compact sources in addition to the classes FRI and FRII of \citet{FR_1974}. 
For the class FRI, the maximum of the radio emission is situated close to the centre of the source. The maxima of the radio emission are located at the edges of the jets for FRII sources. Unresolved and point sources are contained in the Compact class. The Bent class consists of sources for which the angle between the jets differs significantly from 180 degrees. The two subtypes narrow-angle tail (NAT) and wide-angle tail (WAT) are further discriminated by the angle. For an illustration of the considered classes (FRI, FRII, Compact and Bent) see Fig.~\ref{fig:my_label}.

\citet{Aniyan_2017,Alhassan_2018,Tang_2019,Samudre_2021} use convolutional neural networks (CNNs) trained on data from the FIRST survey \cite{FIRST_1995} for the classification of radio galaxies. 

\section{Data}

We combine different catalogues that characterise radio sources from the FIRST survey  to create a data set of radio galaxy images with morphological labels \cite{Gendre_2008,Gendre_2010,Capetti_2017a,Capetti_2017b,Baldi_2017,Proctor_2011, Miraghaei_2017}. 
By explicitly comparing the coordinates of the sources, 300 duplicates were found and removed. Moreover, we exclude 147 sources from the data set that appear in different catalogues with different labels. The resulting data sets are presented in Table~\ref{tab:data set}.
We adopt the preprocessing from \cite{Aniyan_2017} after cropping the images to the input size of our generative network (128 x 128 pixels). In particular, we set all pixel values below three times the local RMS noise to the value of this threshold. Subsequently, the pixel values are rescaled to the range between -1 and 1 to represent floating point greyscale images. We apply classical augmentation to each image during training.

\begin{table}
    \centering
    \begin{tabular}{c|c|c|c|c|c}
         \hline
          & \text{FRI} & \text{FRII} & \text{Compact} & \text{Bent} & \text{Total}  \\ \hline 
         \text{train} & 395 & 824 & 291 & 248 & 1758 \\ 
         \text{validation} & 50 & 50 & 50 & 50 & 200 \\ 
         \text{test} & 50 & 50 & 50 & 50 & 200 \\ 
         \hline 
         \text{total} & 495 & 924 & 391 & 348 & 2158 \\ \hline
    \end{tabular}
    \caption{Number of sources per class in the data sets.}
    \label{tab:data set}
\end{table}

\section{Wasserstein GAN}

Generative adversarial networks (GANs) \cite{goodfellow2014generative,salimans2016improved} are able to learn a representation of the underlying statistical distributions of sets of images. Sampling from those representations may provide additional data points for subsequent treatments \cite{Buhmann_2021,Hadrons_2021}.  

For this project, we employ a variant of the standard GAN setup called Wasserstein GAN (wGAN), which uses the Wasserstein-1 metric, often referred to as the Earth Mover's distance, as main term in the loss function ~\cite{arjovsky2017wasserstein}. A direct advantage of this setup is the correlation between image quality and the value of the loss function, transforming the discriminant of a standard GAN into a critic. Additionally, training of wGANs is often more stable and more likely to converge than standard GAN setups. To approximate the Wasserstein-1 metric by use of a critic network, it has to be ensured that the 1-Lipschitz constraint is fulfilled. This is achieved by applying a gradient penalty term to the loss function as in~\cite{gulrajani2017improved}.

\begin{figure}
    \centering
    \vspace{0.2cm}
    \includegraphics[width=0.8\textwidth,trim={5cm 5cm 5cm 5cm},clip]{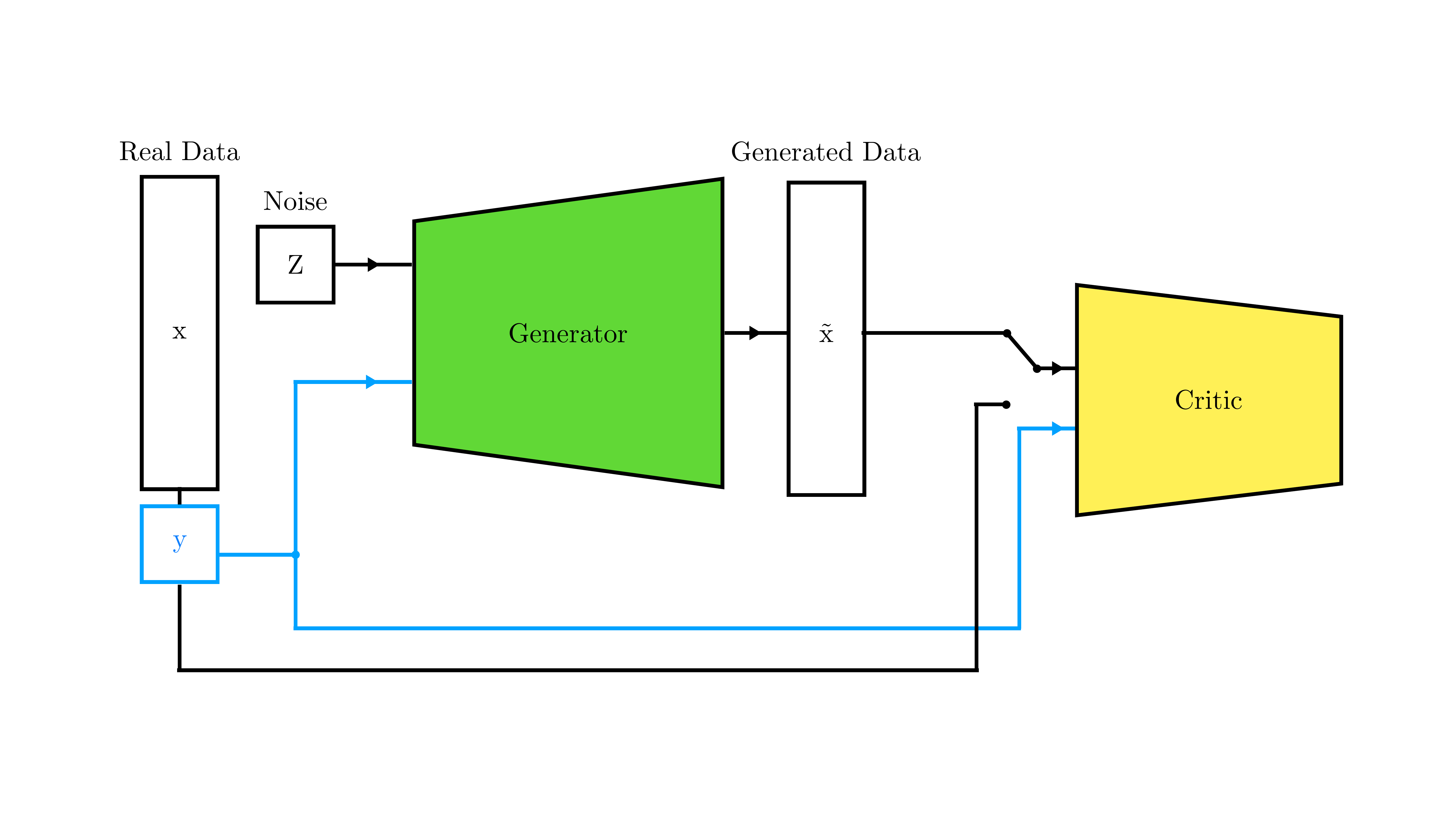}
    \caption{Schematic of the wGAN architecture, where $x$ denotes real images and $y$ the class label. The noise vector $Z$ is the input of the generator and the output of the generator $\tilde{x}$ represents generated images. The critc receives either real or generated images as input.}
    \label{fig:wgan_schematic}
\end{figure}

Since different morphologies of radio galaxies result in very different images, it seems reasonable to condition the networks with the class label. For this setup, this is achieved for the generator by applying a 2D transposed convolution operator on a matrix of image dimensions filled with the class label. The transpose-convoluted layer is then concatenated to the first transpose-convoluted layer of the noise tensor. Batch normalisation in 2D and ReLU activation functions are used. The concatenated tensor is then passed through five additional 2D transposed convolutions, where no normalisation or activation is applied after the last layer. Instead, the individual pixel values are simply clipped to [-1,1] for easy conversion to greyscale.
The critic is built analogously, but uses 2D convolutional layers resulting in a single output node representing the critic score for image quality. Here, layer normalisation and leaky ReLU functions are used except for the last layer. A schematic of the wGAN setup can be found in Fig.~\ref{fig:wgan_schematic}.

\section{Generated Images}

\subsection{Training}

To evaluate statistical fluctuations, ten statistically independent wGAN training runs are launched solely on the training set. The training is performed with a single NVIDIA A100 GPU provided by the Maxwell cluster at DESY. The training is done for 40,000 generator iterations, while the critic is trained five times per generator iteration. A batch size of 400 is chosen and the full training takes roughly 7 hours to complete. The generator and critic weights are saved every 250 iterations, allowing to scan for the best training state later on.

\subsection{Image Quality}

\begin{figure}
    \centering
    \includegraphics[width=0.8\textwidth]{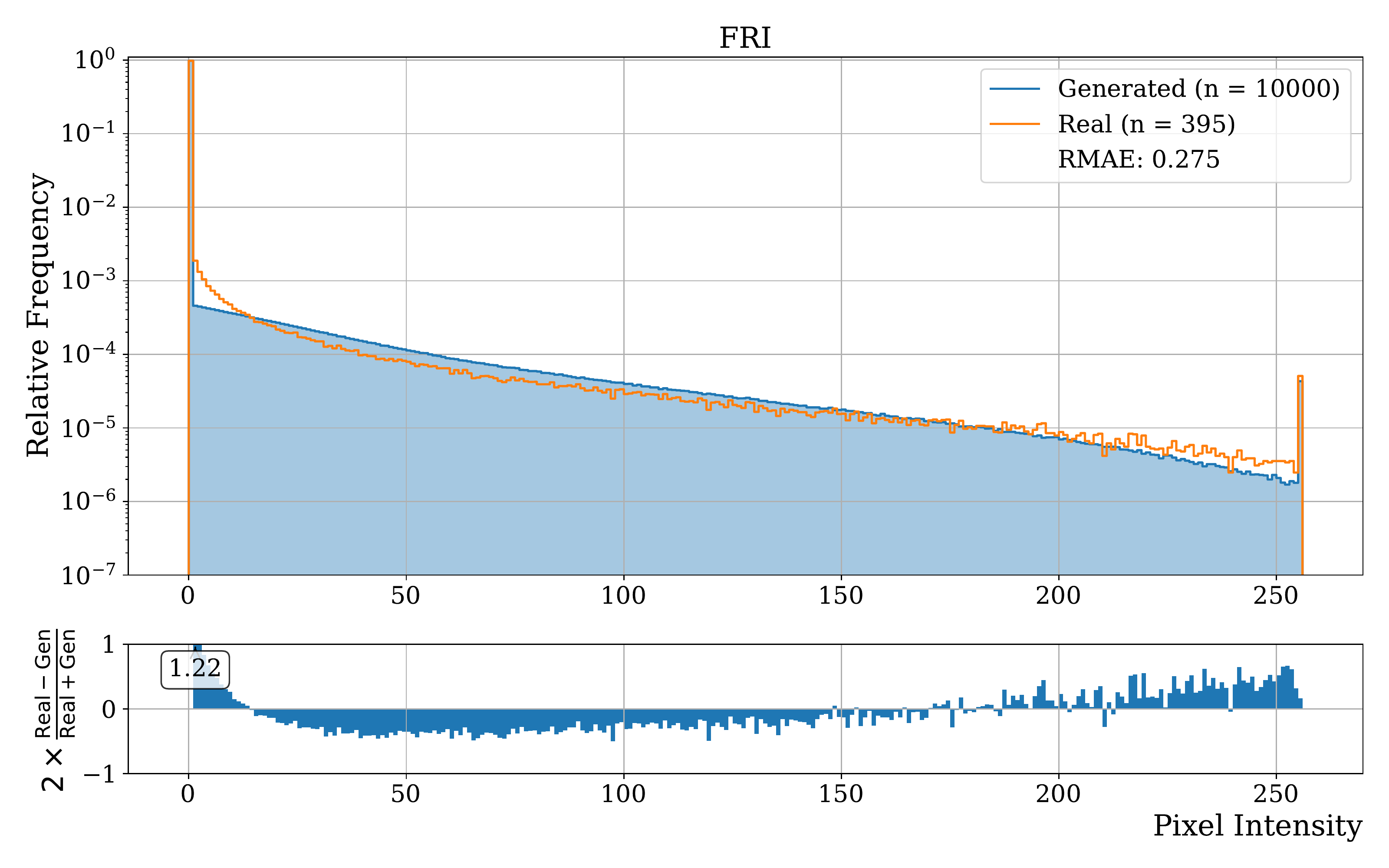}
    \caption{Histogram of pixel intensities of FRI sources comparing batches of real (orange) and generated images (blue) for the training epoch with the lowest combined RMAE. The per-bin relative error is shown in the bottom panel.}
    \label{fig:histogram_example}
\end{figure}

In order to determine the quality of images and thus to find the best performing training iteration, a set of distributions is defined to compare generated images to the validation data set. This includes normalised histograms of pixel intensities, the number of pixels with an intensity greater than zero and of the sum of intensities. These histograms are compared for each class and the relative mean absolute error (RMAE) between the generated set of 10,000 images and the validation set is computed. The RMAE for the different histograms are summed up to yield a single figure-of-merit (FOM) per class, where the training iteration with the lowest FOM value is used in the following. As an example, the histogram of pixel intensities for FRI is depicted in Fig.~\ref{fig:histogram_example}, where the generated histogram is shown in blue and the validation set in orange.

\begin{figure}
    \centering
    \vspace{0.2cm}
    \includegraphics[width=0.9\textwidth,trim={3cm 4cm 3cm 4cm},clip 
]{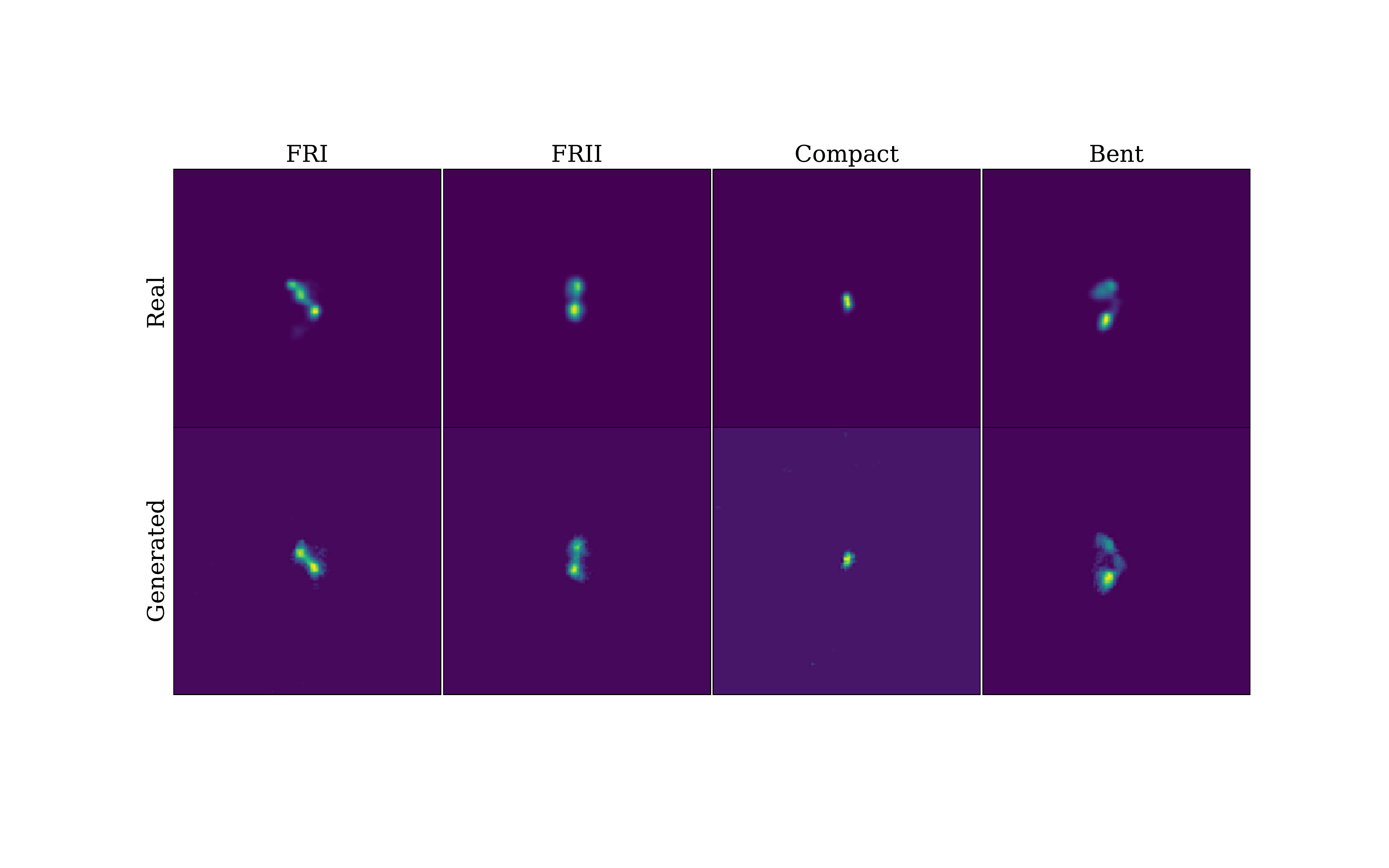}
    \caption{Closest matching pairs of generated and real images.}
    \label{fig:image_examples}
\end{figure}

For a direct visual comparison of image quality, a set of 5,000 images per class is generated and aligned using principle component analysis. Subsequently, the pixel-by-pixel difference is computed for all possible pairs of real and generated images. The closest pairs with a minimal activity for each class are shown in Fig.~\ref{fig:image_examples}.



\section{Using wGAN-supported Augmentation}

As described above, classical augmentation relies on rotation and reflection of training images, adding no statistical information on the morphology. This type of augmentation is, thus, mainly a way to help a classifier understand the rotational invariant nature of radio galaxy images. Nowadays, this could be achieved by directly using equivariant models that incorporate such symmetry constraints inherently and show potential to improve classifier training without relying exclusively on augmentation \cite{Bowles_2021}. 

Here, we evaluate the augmentation achieved by using a combined data set of real and generated images (i.e. wGAN-supported augmentation) by comparing it to using the classically augmented (real-only) data set. The combined data set is created before starting the classifier training with a fixed ratio between the number of generated $i_g$ and real images $i_r$, denoted $\lambda = i_g/i_r$. We study the range $\lambda=1,...,4$ and the class distribution of the generated images is designed such that the combined data set is balanced. For the real-only training runs, class weights are introduced to the loss function to account for the class imbalance in the data set.

We train a simple classifier, namely a fully-connected neural network with three hidden layers and four output nodes. The classifier is trained for 12,500 epochs with a batch size of 100 and a learning rate of $5\cdot10^{-4}$. Training takes on average 3.5 seconds per epoch on an NVIDIA V100 GPU for real-only runs. This setup is not optimised to reach maximal classification accuracies as the goal of this study is only to compare classical with wGAN-supported augmentation.

\begin{figure}
    \centering
    \includegraphics[width=0.8\textwidth]{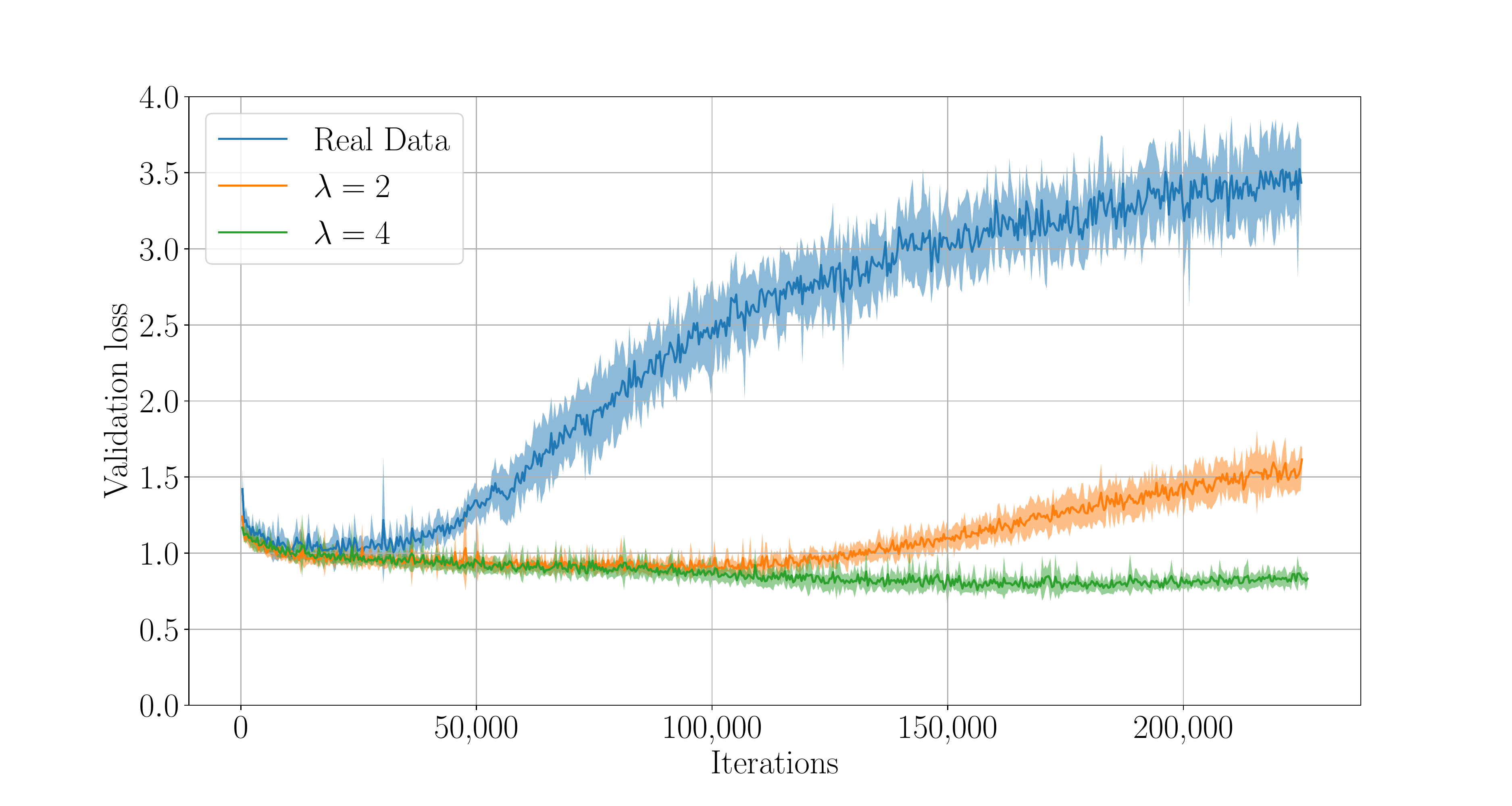}
    \caption{Validation loss during training of the classifier for different values of $\lambda$. The solid line represents the mean of ten independent training runs and the shaded area represents the symmetrised standard deviation.}
    \label{fig:GAN_augmentation} 
\end{figure}

We observe a significant improvement when increasing the number of generated images in the combined data set. As shown in Fig.~\ref{fig:GAN_augmentation} 
the real-only training run enters into over-training (i.e. the value of the loss function increases, while the training accuracy deviates further from the validation accuracy) much earlier than training runs using a combined training data set. The start of the over-trained period depends on $\lambda$, where we observe that adding more generated images to the data set shifts the start to later training iterations, while the validation loss score continues to decrease and the validation accuracy score (not shown) increases.

\begin{figure}
    \centering
    \includegraphics[width=0.8\textwidth]{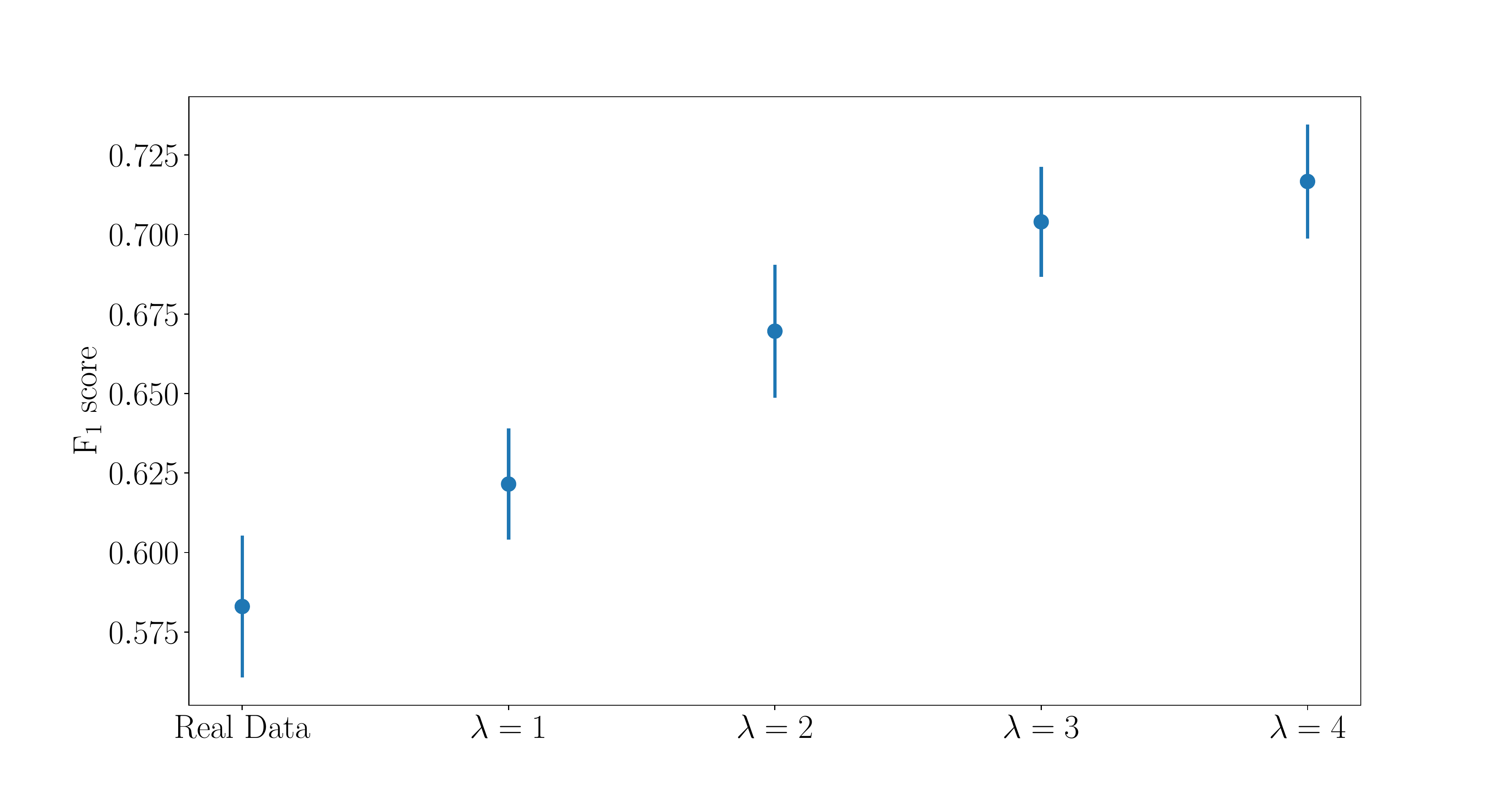}
    \caption{Performance of the best model on an independent test set. We show the mean of ten independent training runs, where the uncertainty is given as the symmetrised standard deviation.}
    \label{fig:F1_GAN}
\end{figure}

In order to compare the overall performance between different training setups, the multi-class Brier score~\cite{Brier_1950}
is used to determine the best training iteration for each of the ten statistically independent training runs for our five different training setups. The Brier score is essentially the mean squared error of the predicted probabilities of a classifier for all classes. This has the advantage that also the certainty of the classifier's decision is considered, which winner-takes-all FOMs such as accuracy do not take into account.
The performance of the best models (i.e. models with minimal Brier score) is evaluated on an independent test set that contains real data only using commonly applied metrics such as accuracy, precision, recall, and $\text{F}_1$ score. The $\text{F}_1$ score is the harmonic mean between precision and recall.
The results for the $\text{F}_1$ score are depicted in Fig.~\ref{fig:F1_GAN}, indicating that training runs using the combined data set with $\lambda=4$ achieve an $\text{F}_1$ score that is \SI[parse-numbers=false]{(23\pm2)}{\percent} higher than for the real-only training runs. 

\section{Discussion}

Previous approaches employing generative models to simulate images of radio galaxies are based on variational autoencoders (VAE) \cite{Ma_8451231,Ma_9023752,Bastien_2021}. Therefore our GAN-based approach is novel in the field of radio astronomy. 
In this work, we demonstrate that we are able to generate highly realistic, artificial images of radio sources. The quality of the generated images is evaluated in different ways and with different metrics. After visual inspection of a large number of generated images, our generated images do not suffer from issues known from other state-of-the-art simulations based on generative networks. In summary, we find:

\begin{itemize}
    \item The resolution of the generated images is as good as the resolution of the images we train on.
    \item The generated images have a similar noise level as the preprocessed training examples.
    \item We do not observe any pseudo-textures or pseudo-structures in the generated images.
 \end{itemize}
 
The images generated with the wGAN can be put to use for applications such as classifier training.
As a result, our contribution represents a major improvement for simulations of radio galaxies based on generative networks. \citet{GANplify_2021} showed for toy models that an amplification of training statistics with generative models is possible. Our work represents an example of this GANplyfication with real astrophysical data. For a recent particle physics application see \cite{Calomplification_2022}.

The presented method allows to amplify the performance of a simple classification model substantially by an increase in the $\text{F}_1$ score of $\SI{23}{\percent}$ for a ratio of $\lambda=4$ between generated and real images and can be particularly useful for applications with unbalanced data sets. 
The application of wGAN-supported augmentation to more complex classifiers such as CNNs is left for future work.

\section*{Acknowledgements}
This work was supported by UHH, DESY, TUHH and HamburgX grant LFF-HHX-03 to the Center for Data and Computing in Natural Sciences (CDCS) from the Hamburg Ministry of Science, Research, Equalities and Districts. This project benefits greatly from the exchange with particle physicists with a vast experience in using generative models for calorimeter simulations and was supported in part through the Maxwell computational resources operated at DESY. 

\printbibliography

\end{document}